\documentstyle[epsf,rotate]{mn}

\newcommand{\ltsimeq}{\raisebox{-0.6ex}{$\,\stackrel 
        {\raisebox{-.2ex}{$\textstyle <$}}{\sim}\,$}} 

\title[On the outburst amplitude of the soft X-ray transients]
{On the outburst amplitude of the soft X-ray transients}
\author[T.~Shahbaz and E.~Kuulkers]{
T.~Shahbaz and E.~Kuulkers\\
University of Oxford, Department of Physics, Nuclear Physics
Laboratory, Keble Road, Oxford, OX1 3RH, UK }

\begin{document}

\maketitle

\begin{abstract} 
\noindent
We find a strong correlation between the optical outburst amplitude
$\Delta V$ and orbital period $P_{orb}$ for the soft X-ray transient
sources with orbital periods less than 1 day. By fitting the observed
values for 8 X-ray transients we determine an empirical relation that can
be used to predict the orbital period of an X-ray transient given only
its outburst amplitude:

\begin{displaymath}
\Delta V= 14.36 - 7.63\log P_{orb}(h).
\end{displaymath}

\noindent
For periods less than 12 hrs we determine a relation for the
absolute magnitude of the accretion disc during outburst, which then
allows us to estimate the distances to the sources.

\end{abstract}
\begin{keywords}
binaries: close -- accretion, accretion disc -- stars: neutron --
novae, cataclysmic variables -- X-rays: stars
\end{keywords}

\section{Introduction}

The soft X-ray transients (SXTs) are a sub-class of the low-mass X-ray
binaries (LMXBs), in which a Roche-lobe overflowing, main sequence or
subgiant star, typically $< 1 M_{\odot}$, orbits a neutron star or black
hole, with an orbital period of a few hours to a few days. The SXTs
undergo episodic periods of X-ray and optical outbursts, which are
separated by long intervals (several decades) of quiescence.

The most fundamental parameter is the orbital period, which sets the size
and evolutionary state of the system. However, one can only determine the
orbital period of the SXTs when they have subsided into quiescence
(unless the system is eclipsing). It is only then one can undertake a
radial velocity study of the secondary star in order to determine the
orbital period. Even then, some systems (e.g. GS1354--64 and A1524--617)
are just too faint to obtain radial velocity curves with conventional
sized telescopes (van Paradijs \& McClintock 1994).

In this letter we derive an empirical formula that allows us to estimate
the orbital period of a SXT given only the size of the optical outburst
amplitude.

\begin{table*}
\caption{General properties of X-ray Novae}
\begin{center}
\begin{tabular}{llccccccl}
Source & X-ray Nova  & Orbital Period  	& V (mags)   &  V (mags)   & E$_{B-V}$
& Distance & References \\ 
       & Designation & (hrs)		& quiescence & outburst   & (mags) 
& (kpc)	   &    \\ \\
GROJ0422+32	& Nova Per 1992 & 5.09  & 22.4	    & 13.2 & 0.30 & 2.0      & 1--4   \\
A0620--00	& Nova Mon 1975 & 7.75  & 18.3	    & 11.2 & 0.35 & 1.1      & 5--8   \\
EXO0748-676	&		& 3.82  & $>$23	    & 16.9 & 0.42 & 7.6	     & 9--13  \\
GRS1009--45	& Nova Vel 1993 & 6.86  & 21.4-21.9 & 13.8 & 0.20 & 1.5--4.5 & 14--16 \\
GRS1124--68     & Nova Mus 1991 & 10.38 & 20.33	    & 13.5 & 0.29 & 3.0--5.5 & 17--20 \\
4U1456--32 	& Nova Cen 1969 & 15.10 & 18.7	    & 12.8 & 0.10 & 1.2	     & 21--24 \\
H1705-250	& Nova Oph 1977 & 12.55 & 21.5	    & 15.8 & 0.45 &          & 25--27 \\
4U1908+005      &               & 18.97 & 19.2      & 14.8 & 0.35 & 2.5	     & 28--29 \\
GS2000+25       & Nova Vul 1988 & 8.26  & 25.2	    & 16.4 & 1.70 & 2.0	     & 30--34 \\
GROJ1655--40    & Nova Sco 1994 & 62.92 & 17.2      & 14.0 & 1.3  & 3.2      & 35--38 \\
GS2023+338      & Nova Cyg 1989 &155.31 & 18.4      & 12.5 & 1.0  & 2.2--3.7 & 39--42 \\ \\
\end{tabular}
\end{center}

\begin{tabular}{ll}
(1) Castro-Tirado, Pavlenko, Shlyapnikovm, Brandt, Lund, Ortiz 1993   & (2) Zhao 1994 \\
(3) Filippenko, Matheson \& Ho 1995 & (4) Beekman, Shahbaz, Naylor, Charles, Wagner, Martini 1997 \\
(5) Wu, Panek, Holm, Schmitz 1983  & (6) McClintock \& Remillard 1986  \\
(7) Haswell, Robinson, Horne, Stiening, Abbott 1993 & (8) Shahbaz, Naylor \& Charles 1994  \\
(9) Pederson \& Mayor 1985 & (10) Wade, Quintana, Horne, Marsh 1985 \\
(11) Parmar, White, Giommi, Gottwald 1986 & (12) Schoembs \& Zoeschinger 1990 \\
(13) van Paradijs \& White 1995 & (14) Della Valle \& Benetti 1993  \\
(15) Shahbaz, van der Hooft, Charles, Casares, van Paradijs 1996  & (16) Della Valle, Benetti, Cappellaro, Wheeler 1997 \\
(17) Della Valle, Jarvis \& West 1991 \& & (18) Cheng, Horne, Panagia, Shrader, Gilmozzi, Paresce, Lund 1992 \\
(19) Orosz, Charles, McClintock, Remillard 1996 & (20) Shahbaz, Naylor \& Charles 1997 \\
(21) Canizares, McClintock \& Grindlay 1980 & (22) Blair, Raymond, Dupree, Wu, Holm, Swank 1984 \\
(23) Shahbaz, Naylor \& Charles 1993 & (24) McClintock \& Remillard 1990 \\
(25) Griffiths et al., 1978 & (26) Remillard, Orosz, McClintock, Bailyn 1996 \\
(27) Filippenko, Matheson, Leonard, Barth, van Dyk 1997  & (28) Charles et al., 1980 \\
(29) Chevalier \& Ilovaisky 1991 & (30) Chevalier \& Ilovaisky 1990 \\
(31) Callanan \& Charles 1991 & (32) Charles, Kidger, Pavlenko, Prokofieva, Callanan 1991 \\
(33) Filippenko, Matheson \& Barth 1995 & (34) Callanan, Garcia, Filippenko, McClean, Teplitz 1996 \\ 
(35) Bailyn et al., 1995 & (36) Hjellming \& Rupen 1995 \\
(37) Horne et al., 1996 & (38) Orosz \& Bailyn 1997 \\
(39) Wagner, Kreidl, Howell, Collins, Starrfield 1989 & (40) Shahbaz, Ringwald, Bunn, Naylor, Charles, Casares 1994 \\
(41) Casares \& Charles 1994 & (42) Casares, Charles, Naylor, Pavlenko 1993 \\ \\
\end{tabular}
\end{table*}

\section{The outburst amplitude--orbital period relation}

In Table 1 we give the general properties of the SXTs. During quiescence
many of the sources show changes in the mean brightness from year to
year, combined with the fact that in some the optical nova was not
discovered optically until several days after the initial X-ray outburst.
We therefore estimate typical errors for the outburst amplitude to be
around 0.2 magnitudes. In Fig. 1 we show the relation between orbital
period and the amplitude of the optical outburst. There is only a lower
limit to the quiescent magnitude of EXO0748-676, therefore it is not
included in Fig. 1. We find that as the orbital period increases, the
amplitude of the outburst decreases. Such a correlation is not
unexpected, as this reflects the fact that at longer orbital periods, the
secondary star is larger and hence more luminous than a system at a
shorter orbital period. The only systems that do not fit this correlation
are the extremely long period systems GROJ1655--40 and GS2023+338; which
both have periods greater than 1 day. Both systems have secondary stars
sufficiently evolved from the main sequence that one cannot compare them
to the other transients. Therefore in what follows we restrict ourselves
to those transients with orbital periods less than 1 day.

For $P_{orb}(h)$ \ltsimeq 24 we find that the $\Delta V$--$\log
P_{orb}(h)$ relationship is well represented by a linear least-squares fit of
the form

\begin{equation}
\Delta V= 14.36(\pm0.78) - 7.63(\pm0.75)\log P_{orb}(h)
\end{equation}

\noindent
which has a correlation coefficient of $-$0.93 (The 1-$\sigma$ errors were
calculated after rescaling the error bars to give a fit with 
$\chi^{2}_{\nu}$ of 1).

Using equation (1) it is therefore possible to estimate the orbital
period of a transient given only the observed values for its outburst
amplitude.

\section{Application}

For some of the X-ray transients, their quiescent magnitudes have proved
too faint for one to obtain accurate radial velocity or photometric light
curves of the secondary star, which will allow one to determine the
orbital period of these systems. In the next section we estimate
quiescent magnitudes and/or the orbital periods of a few of the faint
transients.

\subsection{EXO0748--676}

For a Roche-lobe filling star the average density, $\rho$ is simply a
function of the orbital period (Frank, King \& Raine 1985). For
EXO0748--676 with $P_{orb}$=3.82 hrs, $\rho$=7.54 g~cm$^{-3}$, which
implies a M3--4V star (Allen 1981). This can be compared with the
spectral type of the secondary star obtained by determining its absolute
magnitude. Using equation (4) we find $M_{v}(2)$=10.2 mags, which implies
an M2--3 V star (Allen 1981). 

Using equation (1) we can also predict the quiescent magnitude of
EXO0748-676. With $V_{O}$=16.9, we find $V_{Q}\sim$26.8 mags, which is
consistent with the observed limit of $V_{Q}>$23 mags (see Table 1).
However, if the secondary star is sufficently evolved for it to be
degenerate, then it would not obey equation (4) and would probably be
fainter. Note that the models of King, Kolb \& Burderi (1996) show that
neutron-star SXTs may require evolved companions, even at short orbital
periods.

\subsection{GRS1354--64(=Nova Cen 1967)}

We can estimate the orbital period for GRS1354--64 by using equation (1).
Since there are no $V$-band measurements for the system in quiescence, we
use the colors of a K0--M0 star in order to determine $V_{Q}$. Using
$R_{Q}$=20.3 mags (Martin 1996), $E_{B-V}$=1.0 mags (van Paradijs \&
McClintock 1994) and (V-R)=0.6--1.2 mags (K0--M0$\sc v/iii$; Allen 1981),
we estimate $V_{Q}$=21.6--22.2 mags. Then using equation (1) with
$V_{O}$=16.9 mags (Pederson, Ilovaisky \& van der Klis 1987) gives
$\Delta V$=4.7--5.3 mags and hence $P_{orb}$=15.4--18.5 hrs, which is
consistent with that obtained by Martin (1996); 15.6 hrs.

\subsection{GRS1716--249(=Nova Oph 1993)}

Masetti et al. (1996) interpret a 14.7 hr periodicity as the superhump
period. If this interpretation is correct, then the true orbital period
should be a few percent shorter. Using equation (1) with $V_{O}$=16.3
mags and $V_{Q}>$21 mags (Della Valle, Mirabel \& Rodriguez 1994) we
estimate $P_{orb}<$18.5 hrs, which is consistent with the superhump
period.

\section{The absolute magnitude of the accretion disc}

The amplitude of the optical outburst, $\Delta$ is the difference in
magnitudes between the system in quiescence and in outburst,

\begin{equation}
\Delta V=V_{Q}-V_{O}=M_{Q}-M_{O},
\end{equation}

\noindent
where $V_{Q}$ and $M_{Q}$ are the apparent and absolute magnitudes of the
system in quiescence, respectively, and $V_{O}$ and $M_{O}$ are the
apparent and absolute magnitudes of the system in outburst, respectively.

In outburst the hot accretion disc dominates the optical flux
[$m_{v}(disc)$], whereas in quiescence the observed optical flux arises
from the secondary star [$m_{v}(2)$] and the contribution from the cool
accretion disc. The magnitude of the system in quiescence $V_{Q}$ is
given by $m_{v}(2)+2.5\log f$, where $f$ is the fraction of light arising
from the secondary star; $f$=1.0 implies all the optical flux in
quiescence come from the secondary stars. Typically $f$ is about 0.5,
i.e. 50\% of the light comes from the secondary star (Chevalier \&
Ilovaisky 1989; McClintock \& Remillard 1986; and Charles 1996). We can
therefore write

\begin{eqnarray}
\begin{tabular}{lcl}
$\Delta V$  & = & $m_{v}(2)+2.5\log f - m_{v}(disc)$  \\
            & = & $M_{v}(2)+2.5\log f - M_{v}(disc)$.
\end{tabular}
\end{eqnarray}

\noindent
Warner (1987, 1995) finds that the luminosity of the companion stars in
cataclysmic variables ($P_{orb}$\ltsimeq 10) are indistinguishable from
main sequence stars. The secondary stars in LMXBs with orbital periods
\ltsimeq 12 hrs will also lie on the main sequence or the terminal age
main sequence (Shahbaz, Naylor \& Charles 1997), and so we extend
Warner's relation to 12 hrs. He finds that the absolute magnitude of the
secondary star can be represented by

\begin{equation}
M_{v}(2) = 16.7 - 11.1\log P_{orb}(h).
\end{equation}

\noindent
Unfortunately, a similar relationship does not exist for evolved stars,
therefore the following is only applicable for systems with un-evolved
secondaries, i.e. systems with $P_{orb}$\ltsimeq 12 hrs.

Since we have relationships for $\Delta V$ and $M_{v}(2)$ as functions of
the orbital period ($P_{orb}(h)$\ltsimeq 12), we can then derive a
similar relationship for $M_{v}(disc)$

\begin{eqnarray}
\begin{tabular}{lcl}
$M_{v}(disc)$ & =  & $M_{v}(2) - \Delta V + 2.5\log f$ \\
              & =  & $2.34(\pm0.78) - 3.47(\pm0.75)\log P_{orb}(h)$ \\
              & $+$ & $2.5\log f.$
\end{tabular}
\end{eqnarray}

As the orbital period increases, the size of the system also increases.
If we assume that during outburst the size of the accretion discs in SXTs
are similar i.e. the accretion discs extends out to the tidal radius,
then one expects the accretion disc during outburst to brighten as the
orbital period of the system increases, simply because of the increase in
the projected surface area of the accretion disc.

Equation (5) can be compared with the formula give by van Paradijs \&
McClintock (1994). They determine a relationship between the absolute
magnitude of the accretion disc, X-ray luminosity and orbital period for
SXTs in outburst and LMXBs. We can rewrite their equation as

\begin{equation}
M_{v}(disc) = 1.57 - 1.51\log P_{orb}(h) - 1.14\log (L_{X}/L_{Edd})
\end{equation}

\noindent
where $L_{X}$ is the outburst X-ray luminosity and $L_{Edd}$ is the
Eddington limited luminosity for a 1.4 M$_{\odot}$ neutron star. (It
should be noted that, although van Paradijs \& McClintock included the
black hole candidates A0620--00 and GS2023+338, removing these points does not
change the correlation significantly.)
Chen, Shrader \& Livio (1997) tabulate $\log(L_{X}/L_{Edd})$ for all the
LMXBs and SXTs in outburst, where in this case $L_{Edd}$ is the Eddington
limiting for an object with general mass $M$. By fitting the data points
for systems with orbital periods less than 12 hrs, we obtain a linear
least-squares fit of the form 

\begin{equation}
\log (L_{X}/L_{Edd}) = 3.63(\pm0.90)\log P_{orb}(h) - 4.20(\pm1.02).
\end{equation}

\noindent
From equations (6) and (7) we obtain the absolute magnitude of the
accretion disc as a function of orbital period:

\begin{equation}
M_{v}(disc) = 6.36(\pm 1.36) - 5.65(\pm1.20)\log P_{orb}(h).
\end{equation}

\noindent
As one can see the gradients of equation (5) and (8) are comparable (at
the 90 per cent confidence level). However, unlike the correlation
derived by van Paradijs \& McClintock (1994), the relation we obtain for
the absolute magnitude of the accretion disc does not depend on the
distance and reddening to the SXT. We therefore believe that equation (5)
is a better representation of the absolute magnitude of the accretion
disc in LMXB for systems with $P_{orb}(h)$\ltsimeq 12.

It is interesting to compare the absolute magnitudes of the accretion
discs in dwarf novae and SXTs in outburst. By manipulating equation (13)
of Warner (1987) for the absolute magnitude of accretion discs in dwarf
novae at maximum light, we find

\begin{equation}
M_{v}(disc,W\!D) = 6.0 - 2.69\log P_{orb}(h).
\end{equation}

\noindent
Comparing this with equation (5), we find for a given orbital period, the
accretion discs in SXTs during outburst are more than 4 magnitudes
brighter (depending on the value for $f$) than dwarf novae at maximum
light, which is typically what is expected (van Paradijs \& McClintock
1994). We note that this is similar to the difference of the mean
absolute magnitudes derived for persistent LMXBs and CVs (e.g.
nova-likes; see van Paradijs \& Verbunt 1984). This difference in
brightness has a natural explanation in that accretion discs of SXTs in
outburst and persistent LMXBs are dominated by X-ray irradiation, whereas
the discs in CVs are not (see e.g. van Paradijs \& Verbunt 1984 and van
Paradijs \& McClintock 1994). However, it should be noted that the
orbital separation of the SXTs, (most of which are black holes) are
probably a factor 2 larger at a given period, therefore disc area is 4
times larger than in the dwarf novae. This combined with the fact that
the accretion rate of a SXT disc in outburst is an order of magnitude
higher than in a dwarf nova at a given period may partly explain the
difference in accretion disc magnitudes.

\section{The distance--period relation}

We can now use equation (5) along with the distance modulus equation to
obtain an expression for the distance to the X-ray transients
($P_{orb}(h)\ltsimeq$ 12) as a function of orbital period:

\begin{equation} 
\begin{tabular}{lcl}
$5\log D_{kpc}$ & =  & $V_{O} - A_{v} - 2.5\log f$ \\
                &$-$ & $12.34 + 3.47\log P_{orb}(h)$, 
\end{tabular}
\end{equation}

\noindent
where $D_{kpc}$ is the distance in kpc and $A_{v}$ is the reddening.

In Table 2 we estimate the distances for 7 transients using this relation
(we have assumed $f$=1.0, so the distance estimates are lower limits). As
one can see for most of the systems our estimates agree quite well with
the values in the literature (see Table 1). We note, however, that the
main uncertainty in determining the distance is the value used for the
reddening.

\section{Conclusions}

By fitting the outburst amplitudes for 8 X-ray transients we determine an
empirical relation that can be used to predict the orbital period of an
X-ray transient. Also, for periods below less than 12 hrs we determine a
relation for the absolute magnitude of the accretion disc during
outburst, which allows us to estimate the distances to the sources.

\begin{table}
\caption{Distance estimate for transients with $P_{orb}(h)<$ 12 and $f$=1.0}
\begin{center}
\begin{tabular}{lccccc}
Source 		& $P_{orb}$	& V (mags)  & $A_{v}$   & Distance \\
                & (hrs)         & outburst  & (mags)    & (kpc)  \\ \\
EXO0748-676	& 3.82		& 16.9	    & 1.3	& 11.4$\pm$2.0 \\
GROJ0422+32	& 5.09  	& 13.2	    & 0.9	&  3.0$\pm$0.6 \\
GRS1009--45	& 6.86  	& 13.8 	    & 0.6	&  5.7$\pm$1.1 \\
A0620--00	& 7.75  	& 11.2	    & 1.1       &  1.5$\pm$0.3 \\
GS2000+25	& 8.26  	& 16.4	    & 5.32      &  2.4$\pm$0.5 \\ 
GRS1124--68     & 10.38 	& 13.5	    & 0.9	&  5.7$\pm$1.2 \\ \\
\end{tabular}
\end{center}
\end{table}

\section*{Acknowledgements}

We would like to thank Phil Charles for valuable discussions and the
referee, Andrew King for his careful reading of the manuscript. The
figure was plotted using the $\sc ark$ software on the Oxford Starlink
node.

\begin{figure*} 
\rotate[l]{\epsfxsize=500pt \epsfbox[00 00 700 750]{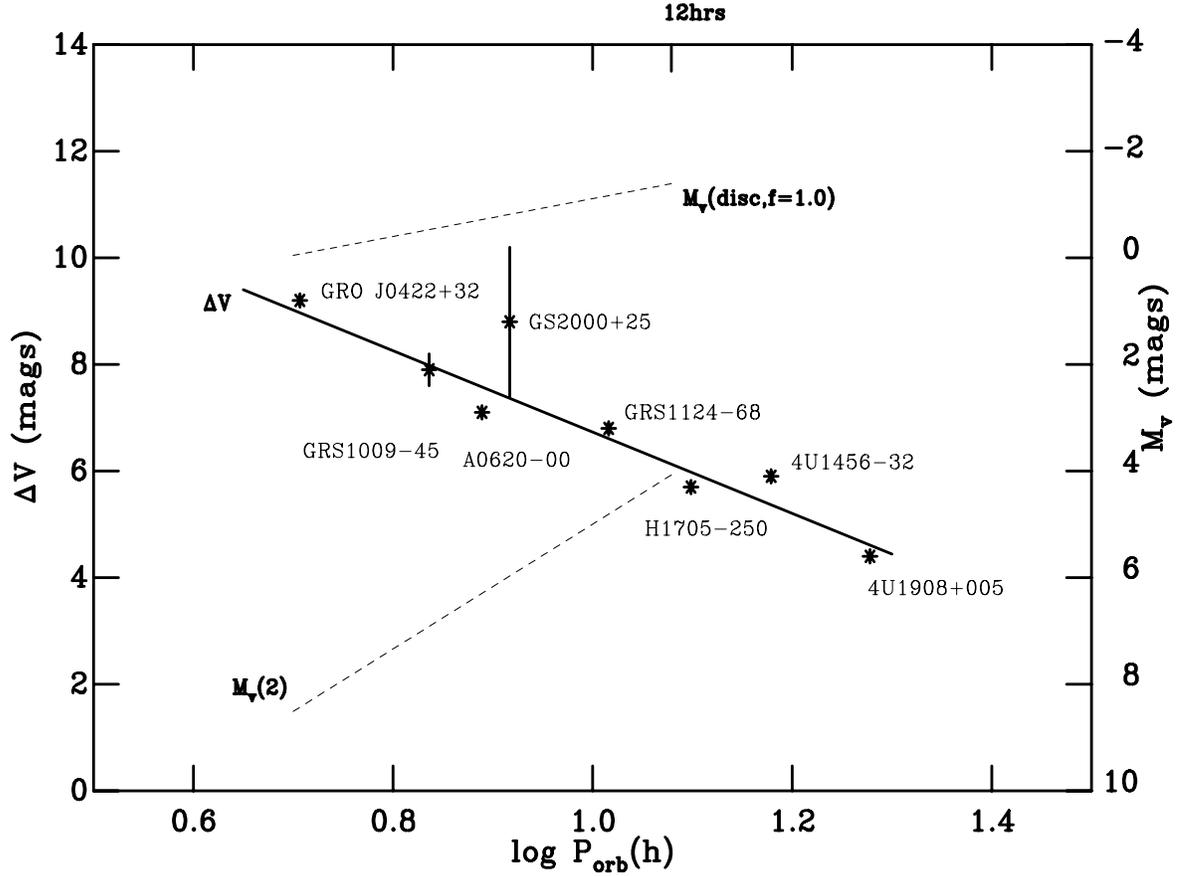}}
\caption{Shown is the $V$-band outburst amplitude versus orbital period for the
soft X-ray transients with orbital periods less than 1 day. For
GRS1009--45 and GS2000+25 the V-band magnitude in quiescence was
estimated using the spectral type of the secondary star; the large
uncertainty is due to the poorly constrained reddening. The solid line is
the least-squares fit to the data points. For systems with periods less
than 12 hrs we also plot (dashed lines) relations for the absolute
magnitude of the accretion disc in an outbursting SXT (assuming the
secondary star dominates in quiescence, i.e. $f=1$) and the absolute
magnitude of the secondary star (Warner 1995). If the secondary star does
not dominate in quiescence the SXT disc is even brighter ($f<1$). The
outburst amplitude versus orbital period correlation is due to the
secondary star brightening faster than the accretion disc, as the orbital
period increases.}

\end{figure*}

\end{document}